\documentclass[3p,preprint,hyperref]{elsarticle}
\input{SMEFTdev_aux}

\begin{document}

\begin{frontmatter}
\title{SMEFT deviations}

%\author[\csuma]{Federico Camponovo}
%\author[\csuma,\csumb]{Giampiero Passarino}

\author[torino]{Federico Camponovo, Giampiero Passarino}
\ead{federico.camponovo@unito.it, giampiero@to.infn.it}

\address[torino]{\csumb}

%\fntext[support]{\support}

%\email{federico.camponovo@unito.it}
%\email{giampiero@to.infn.it}

\begin{abstract}
This work is based on a bottom{-}up approach to the standard{-}model
effective field theory (SMEFT), resulting in an equiprobable space of
Wilson coefficients. The randomly generated Wilson coefficients of the
SMEFT (in the Warsaw basis) are treated as pseudo-data and, for each observable, the
corresponding probability density function is computed. The goal has been
to understand how large are the deviations from the SM once the SMEFT scale 
($\Lambda$) and the range of the Wilson coefficients are selected. Correlations 
between different observables are also discussed.
%---
\end{abstract}
%--
\begin{keyword}
Effective Field Theories
\PACS 11.10.Gh \sep 11.10.Lm \sep 11.15.Bt
\end{keyword}

\end{frontmatter}

%--
\section{Introduction \label{Intro}}
%--

The standard{-}model effective field theory~\cite{Manohar:2018aog} (SMEFT) is a useful tool to
analyze possible deviations from the standard model (SM). In this work we
will use the SMEFT (in the so{-}called Warsaw basis~\cite{Grzadkowski:2010es}) 
at the one{-} loop level. 
This means insertion of $\mrdim = 6$ operators in one{-}loop SM diagrams plus 
``pure" one loop SMEFT diagrams (\ie diagrams with no counterpart in the SM). 
In particular we will study the following (pseudo-)observables:

\begin{itemize}
    \item[\snitem] the $\mathrm{g} - 2$ of the muon, \ie the anomalous magnetic moment $\mra_{\PGm}$,
    \item[\snitem] the value for the $\PW$ boson mass, $\mrM_{\mathrm{W}}$, as derived by using
            the LEP1 input parameter set (IPS),
    \item[\snitem] the vector and axial couplings for the $\PZ{-}\,$boson decay
            $\PZ \to \PGmp \PGmm$ and the corresponding $\sin^2\theta^{\PGm}_{\eff}$,
    \item[\snitem] the Higgs boson decay $\PH \to \PGg \PGg$,
    \item[\snitem] the Higgs boson decay $\PH \to \PAQb \PQb$.
\end{itemize}

It is not the goal of this paper to describe the use and reuse of the SMEFT (for that see \Bref{David:2020pzt})
in 
fits~\cite{Brivio:2017btx,Ellis:2018gqa,Murphy:2017omb,Ellis:2020unq,Ethier:2021bye,https://doi.org/10.48550/arxiv.2211.08353}. The goal is instead to 
compute SMEFT deviations w.r.t. the SM and to understand how large they can be and the correlation among 
different (pseudo-) observables.
The strategy of the calculation will be as follows. Given an observable
$\mcO$ we first compute $\mcO^{(4)}$, \ie $\mcO$ at dimension four (the SM). If $\mrM_{\mcO}$ is
the corresponding matrix element then

\begin{equation}
\mcO^{(4)} = \int \mathrm{dPS}\,\sum_{\spin} \mid \mrM^{(4)}_{\mcO} \mid^2 ,
\end{equation}

where $\mathrm{dPS}$ indicates the corresponding phase{-}space integration.
Including SMEFT (\ie dimension six operators) we will have

\begin{equation}
\mrM_{\mcO} = \mrM^{(4)}_{\mcO} + \frac{\mrg_6}{\sqrt{2}}\,\mrM^{(6)}_{\mcO} , \qquad \mrg_6 = \frac{1}{\mrG_{\mrF}\,\Lambda^2} ,
\label{defg6}
\end{equation}

giving the full $\mcO^{(6)}$. Note that both $\mrdim = 4$ and $\mrdim = 6$ terms may or may not contain
loop corrections. Here we have introduced $\mrg^{-1}_6= \mrG_{\mrF}\,\Lambda^2$, where
$\mrG_{\mrF}$ is the Fermi coupling constant and $\Lambda$ is the SMEFT scale.
Two options will follow, linear or quadratic SMEFT. In the first case only
the linear term will be kept in squaring the amplitude.

After having computed the relevant quantities we will define deviations

\begin{equation}
\Delta\,\mcO = \mcO^{(6)}/\mcO^{(4)} - 1 ,
\end{equation}

and \dnuma plot the relative probability density function (pdf) by randomly generating the corresponding set 
of Wilson coefficients, \dnumb display relationships between observables. 

\begin{itemize}

\item[\snitem] In a top{-}down approach we are assuming a space of UV{-}complete models $\mrM_i(\{\mrp_{\mrM}\})$ 
functions of their  parameters (masses and couplings). 
By UV{-}completion we mean passing from the SM to a more general quantum field theory (QFT) above a threshold; 
the more general QFT should explain more experimental data than the SM. We are not addressing here the question posed by
Georgi~\cite{Georgi:1993mps}: \textit{it may even be possible that there is no end and, simply 
more and more scales as one goes to higher and higher energies}.
From the space $\mcM$ we ideally derive the corresponding low{-}energy limit, taking
into account those models that can be described by the SMEFT; these limits and mixings of heavy Higgs bosons
are discussed in \Bref{Passarino:2019yjx} heavy{-}light contributions in \Bref{delAguila:2010mx}.
In the low{-}energy limit we obtain a space of Wilson coefficients $\mcW(\{\mra\})$; each parameter of $\mcM$ is 
now translated into a set of Wilson coefficients. Of course there will be limitations on $\mcM$, \eg models $\mrM_i$ 
should respect custodial symmetry~\cite{Low:2012rj}.

\item[\snitem] In a purely bottom{-}up approach~\cite{Hartmann:2001zz} (our framework), $\mcW$ defines our computable 
``theory",  see \Bref{Passarino:2021uxa}.

\end{itemize}

In order to arrive at the final result we will have to (briefly) discuss
the main ingredients of the calculation.

\section{SM and SMEFT renormalization \label{Ren}}

We will not discuss renormalization of the SMEFT in great details. To
summarize what has been done (for full details see~\Brefs{Costello2011,Ghezzi:2015vva,Passarino:2019yjx}) we will 
define a renormalization procedure.

\begin{itemize}
\item After the specification of the gauge fixing term including the
corresponding ghost Lagrangian we select a renormalization scheme
and a choice of the IPS. Every counterterm for the parameters in the
Lagrangian contains an arbitrary UV{-}finite constant. Any explicit
definition of the constant is a definition of the renormalization
scheme.

\item It will be enough to say that the SMEFT, as any QFT, depends on
parameters and on fields.  Our strategy is to define counterterms
(CT) for the parameters, to introduce an IPS and to perform
on{-}shell renormalization for the SM parameters and the $\MSB$
renormalization for the Wilson coefficients, see
\hyperref[det]{\ref*{det}}. Beyond two{-}point
functions this will require a mixing of the Wilson coefficients.
\end{itemize}

Some comment is needed on external leg wave{-}function factors:
external unstable particles represent a notorious problem, see~\Brefs{Actis:2006rb,Actis:2006rc}. 
Here we will follow the strategy developed for LEP observables. As a
consequence the wave function factors are taken to be real; of course,
one should introduce the notion of complex poles but the
$\PZ${-}observables (and also the $\PH${-}observables) have not been described
in terms of complex poles.

Removal of UV{-}poles is not the end of any renormalization procedure, see \Bref{Bardin:1999ak}.
For instance, after removing the UV{-}poles, we need to connect the
renormalized $SU(2)$ coupling constant ($\mrg_{\mrR}$) to the fine structure constant $\alpha(0)$ or to
the Fermi coupling constant $\mrG_{\mrF}$.
Most of the renormalization procedure has to do with two{-}point functions, therefore we recall few definitions

\bqa
\mrS_{\PGg\PGg} &=& \frac{\mrg^2\,\stws}{16\,\pi^2}\,\Pi_{\PGg\PGg}(p^2)\,p^2 ,
\qquad
\mrS_{\PZ\PGg} = \frac{\mrg^2\,\stw}{16\,\pi^2\,\ctw}\,\Sigma_{\PZ\PGg}(p^2) ,
\nl
\mrS_{\PZ\PZ} &=& \frac{\mrg^2}{16\,\pi^2\,\ctws}\,\Sigma_{\PZ\PZ}(p^2) ,
\qquad
\mrS_{\PW\PW} = \frac{\mrg^2}{16\,\pi^2}\,\Sigma_{\PW\PW}(p^2) ,
\eqa

where $\stw$ is the (bare) sine of the weak{-}mixing angle. The procedure is as follows:

\begin{enumerate}
\item[\dnuma] Introduce the transition $\mrS_{\PZ\PGg}$ and the corresponding
self{-}energy $\mrS_{\PGg\PGg}$.

\item[\dnumb] Define the $\PGg \PAl \Pl$ vertex at LO (but including SMEFT terms) and
compute the corresponding amplitude for Coulomb scattering obtaining
        
\begin{equation}
\mcA^{\mrc}_{\myLO} = \Gamma_{\myLO}\,\gamma^{\mu} \,\otimes\,\gamma_{\mu} .
\end{equation}

\item[\dnumc] Introduce the Dyson resummed $\PGg$ propagator and define 
(residue of the pole at zero momentum transfer)
            
\begin{equation}
\mcA(\mrg_{\mrR}) = \Gamma^2_{\myLO}\,\Bigl[ 1 - \Pi_{\PGg\PGg}\mid_{p^2=0} \Bigr]^{-1} .
\end{equation}

\item[\dnumd] Introduce CTs for $\mrg$ and mix the Wilson coefficients. It will
follow that the dimension four is UV{-}finite after introducing the
parameter CTs and dimension six is UV{-}finite after mixing.
    
\item[\dnume] Next we write the equations

\bq
\mcA(\mrg_{\mrR}) = 4\,\pi\,\alpha(0) ,
\quad
\mrg^2_{\mrR} = \frac{4 \pi\,\alpha(0)}{\stws}\,\Bigl\{
1 + \frac{\mrg_6}{\sqrt{2}}\,\delta \mrg^{(6)}_0 + \frac{\alpha(0)}{\pi}\,\Bigl[
\delta \mrg^{(4)} + \frac{\mrg_6}{\sqrt{2}}\,\delta \mrg^{(6)}_1 \Bigr] \Bigr\} ,
\eq

where $\mrg_6$ id defined in \hyperref[defg6]{Eq.(\ref*{defg6})},
and fix the $\delta\,\mrg$ (UV{-}finite) CTs.
        
\end{enumerate}

Of course, this procedure requires the well{-}known WST identities for
the cancellation of vertices and wave function factors at $p^2 = 0$.
What we have here is that this WST identity is, well known for QED,
proven for the SM~\cite{Bardin:1999ak,Dittmaier:2021swk}, proven to be valid in 
the SMEFT by our explicit calculations.

Of course we also need the relation between $\mrg_{\mrR}$ and the Fermi coupling
constant. The derivation follows in a similar way.

\section{Computing (pseudo-)Observables \label{details}}
We have selected the IPS containing $\mrM_{\sPW}$ and $\mrM_{\sPZ}$.
The first observable to be considered is the $\mrg - 2$ of the muon. After
computing the (QED) Schwinger term we have included the one{-}loop SM
contributions and added the dimension six contributions. The strategy is,
as described, to make UV{-}renormalization, to separate the IR/collinear
QED corrections, and to develop an efficient algorithm for computing the
$p^2 \to 0$ limit.
For a model{-}independent analysis of the magnetic and electric
dipole moments of the muon and electron see \Bref{Aebischer:2021uvt}.

As far as $\PZ \to \PGmp \PGmm$ is concerned we will adopt the LEP
strategy, see \Brefs{Z-Pole,Bardin:1999gt}: the explicit formulae for the 
$\PZ\PAl\Pl$ vertex are always written starting from a Born-like 
form of a pre-factor times a fermionic current, 
where the Born parameters are promoted to effective, scale-dependent parameters

\begin{equation}
\gamma^{\mu}\,\Bigl( \mathcal{G}^{\Pf}_{\PV} + \mathcal{G}^{\Pf}_{\PA}\,\gamma^5 \Bigr) .
\end{equation}

The corresponding width (LEP1 conventions) is

\bq
\Gamma(\PZ \to \PAl\Pl) = 4\,\Gamma_0\,\Bigl[ 
\mid \mathcal{G}^{\Pf}_{\PV} \mid^2\,\mrR^{\Pl}_{\PV} +
\mid \mathcal{G}^{\Pf}_{\PA} \mid^2\,\mrR^{\Pl}_{\PA} \Bigr] , \qquad
\Gamma_0 = \frac{\mrG_{\mrF}\,\mrM^3_{\sPZ}}{24\,\sqrt{2}\,\pi} ,
\eq

where the radiator factors describe the final state QED and QCD corrections and take into
account the fermion mass.
Next define $\mrg^{\Pf}_{\PV\,,\,\PA}$ as the real part of
$\mathcal{G}^{\Pf}_{\PV\,,\,\PA}$ and define

\begin{equation}
\sin^2\theta^{\mu}_{\eff} = \frac{1}{4}\,\Bigl( 1 -
    \frac{\mrg^{\mu}_{\PV}}{\mrg^{\mu}_{\PA}} \Bigr) .
\end{equation}

The processes $\PH \to \PGg\PGg$ and $\PH \to \PAQb \PQb$ do not need additional comments.

\begin{figure}[t]
   \centering
   \vspace{-3.cm}
\includegraphics[height=7.cm,width=7.cm]{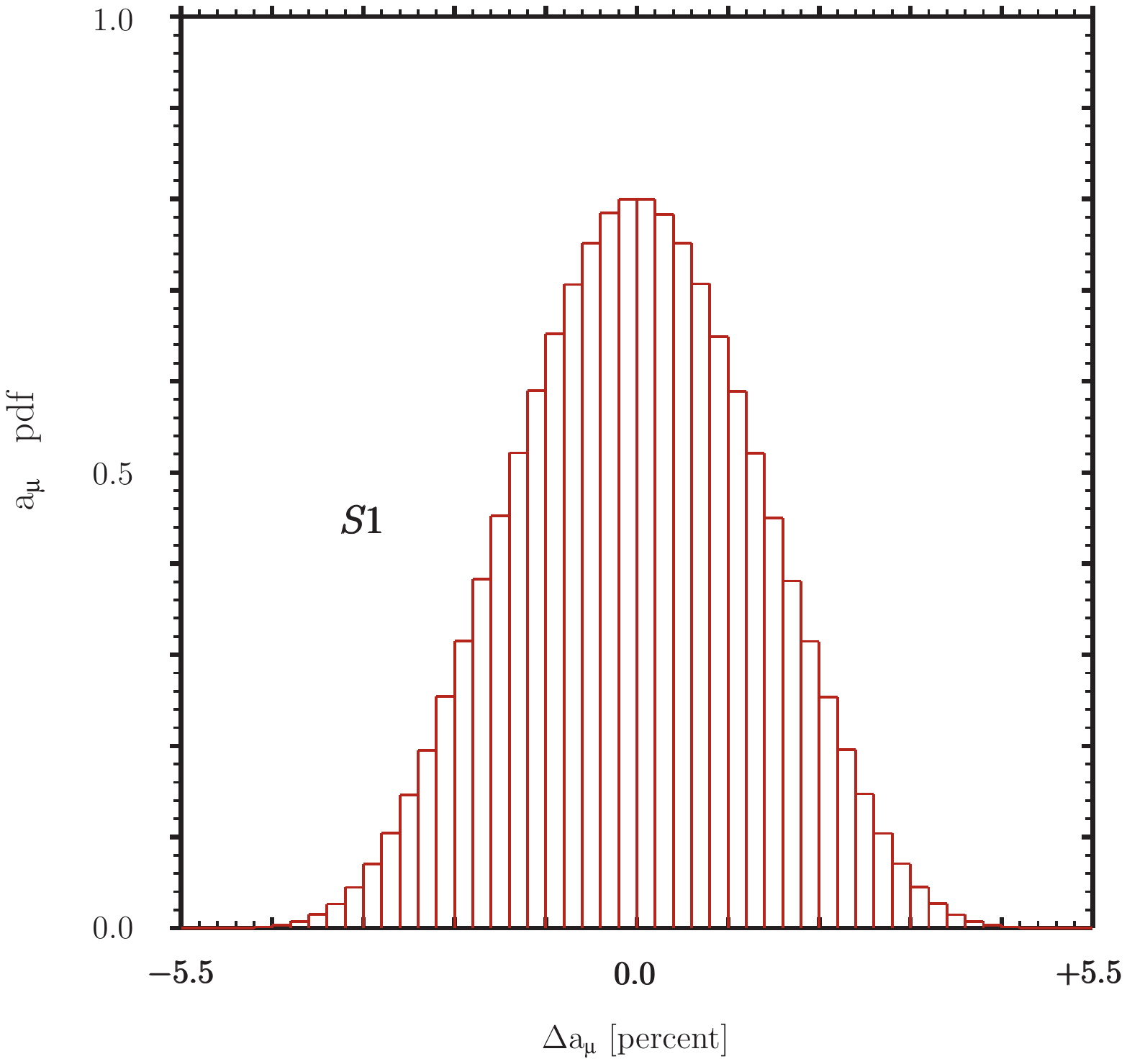}
\includegraphics[height=7.cm,width=7.cm]{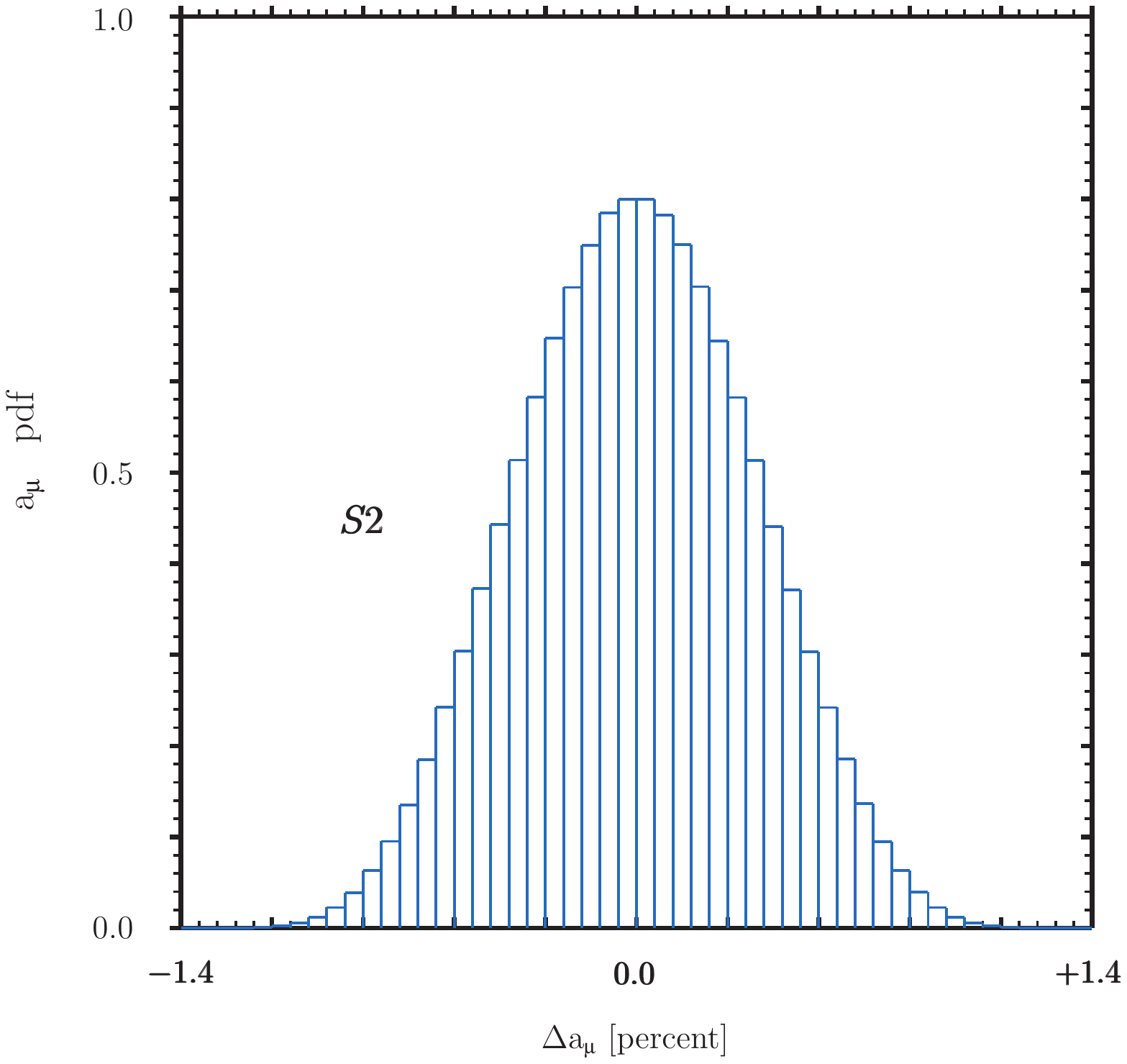}
%\vspace{-1.5cm}
\vspace{-2.cm}
\caption[]{The pdf for $\Delta \mra_{\PGm}$. Left figure refers to scenario S1, right figure to scenario S2.}
\label{devfig1}
\end{figure}

\begin{figure}[t]
   \centering
%   \vspace{-2.cm}
   \vspace{-1.cm}
\includegraphics[height=7.cm,width=7.cm]{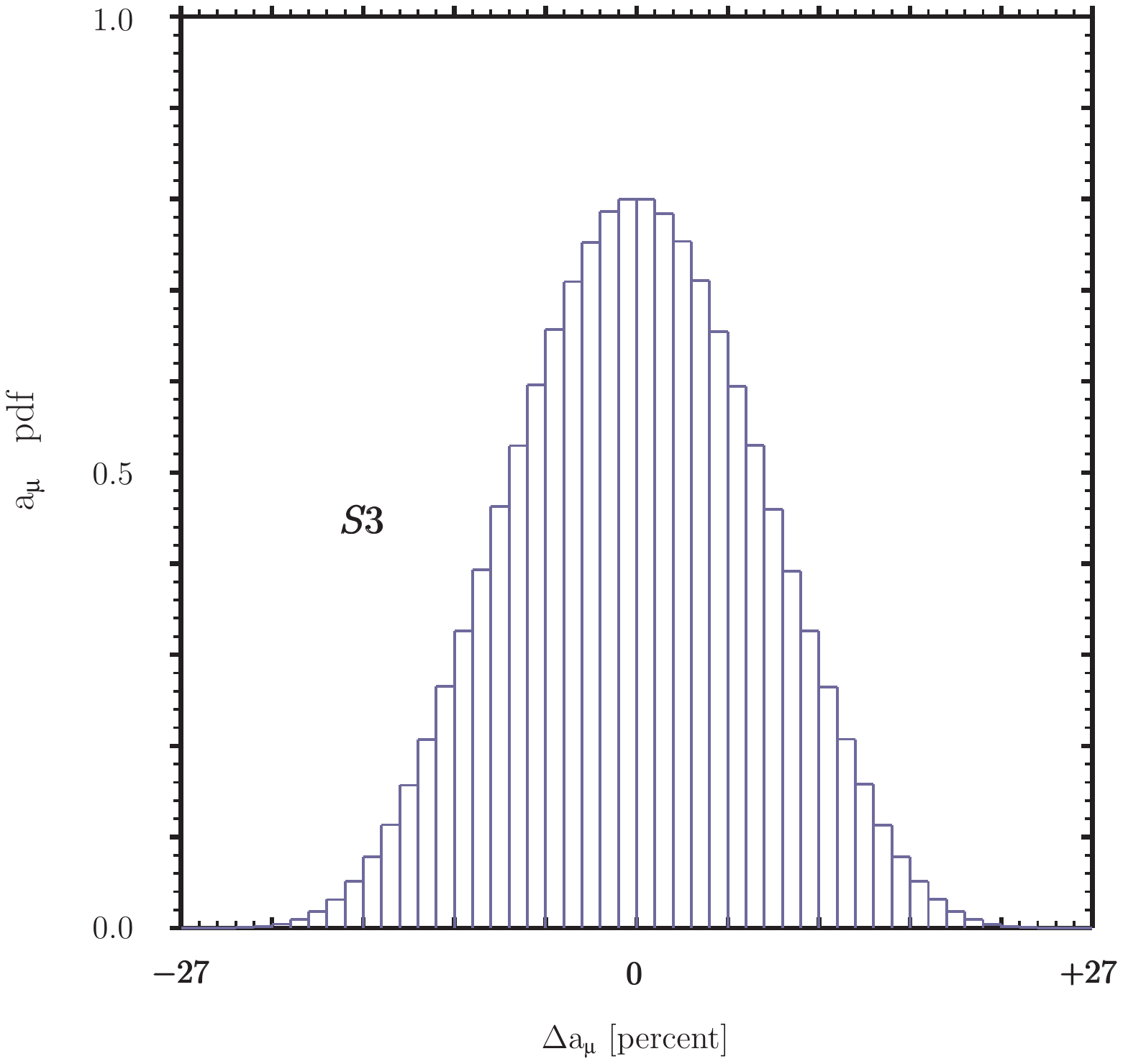}
\includegraphics[height=7.cm,width=7.cm]{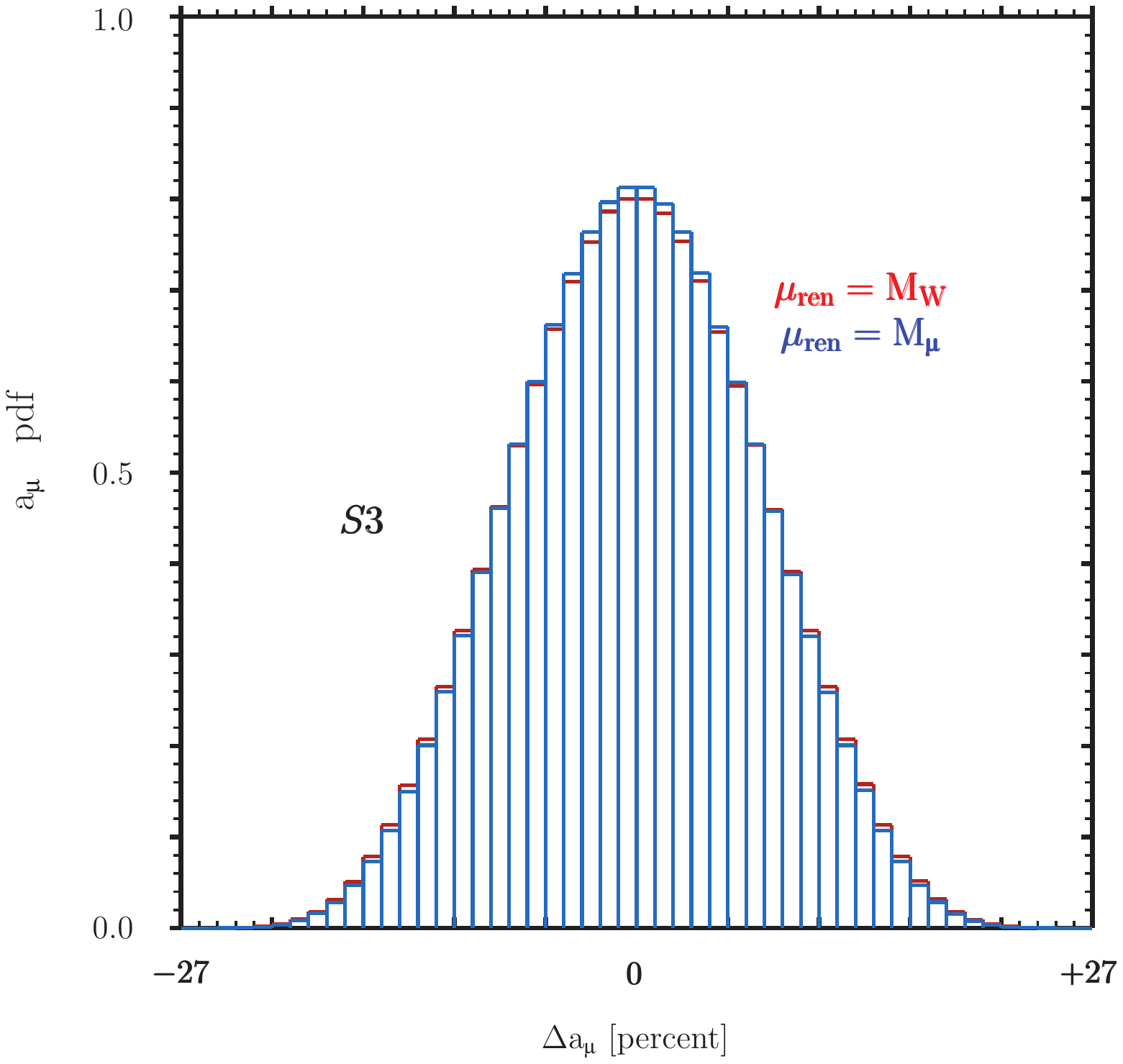}
\vspace{-2.cm}
\caption[]{The pdf for $\Delta \mra_{\PGm}$. $\Lambda = 1\UTeV$ and Wilson coefficients $\in\,[-0.5\,,\,+0.5]$.}
\label{devfig2}
\end{figure}

Predicting the $\PW$ mass means changing the IPS. Therefore we start from the following equations:
%--
\bqa
\frac{1}{\mrg^2 \stws} &=& \frac{1}{4 \pi\,\alpha} + \frac{1}{16\,\pi^2}\,\Pi_{\PGg\PGg}(0) ,
\nl
8\,\frac{\mrM^2}{\mrg^2} &= & \frac{\sqrt{2}}{\mrG_{\mrF}} + \frac{1}{16\,\pi^2}\,
\Bigl[ \Sigma_{\PW\PW}(0) + \mrM^2\,\stws\,\delta_{\mrG} \Bigr] ,
\nl
\frac{\mrM^2}{\ctws} &=& \mrM^2_{\sPZ} + \frac{\mrg^2}{16\,\pi^2\,\ctws}\,\Re\,\Sigma_{\PZ\PZ}(\mrM^2_{\sPZ}) ,
\label{LEP1re}
\eqa

where $\mrg, \stw$ and $\mrM$ are bare parameters and $\Pi$ and $\Sigma$ are self{-}energies containing
dimension six contributions. 
In the second equation $\delta_{\mrG}$ collects the contributions to the muon decay coming from vertices and
boxes. While $\delta^{4)}_{\mrG}$ has been known since a long time~\cite{Hollik:1988ii} we still miss 
a complete calculation of $\delta^{(6)}_{\mrG}$ . 

By solving the set of renormalization equations we obtain bare parameters as
a function of experimental data. It is always convenient to resum large logarithmic corrections so that the
lowest{-}order resummed solution for the weak{-}mixing angle will be

\bq
\stws = {\bar{\mrs}}^2 = \frac{1}{2}\,\Bigl\{
1 - \Bigl[ 1 - 4\,\frac{\pi\,\alpha(\mrM^2_{\sPZ})}{\sqrt{2}\,\mrG_{\mrF}\,\mrM^2_{\sPZ}} \Bigr]^{1/2} \Bigr\} .
\eq

By $\stw$ we mean the Lagrangian parameter while will reserve the notation $\theta_{\sPW}$ for
$\sin^2\theta_{\sPW} = 1 - \mrM^2_{\sPW}/\mrM^2_{\sPZ}$.
Inserting the solutions into the inverse $\PW$ propagator returns $\mrM_{\sPW}$ with a lowest{-}order solution
given by $\mrM_{\sPW} = {\bar{\mrc}}\,\mrM_{\sPZ}$ which receives corrections in perturbation theory, including
the ones due to dimension six operators.

It is worth noting that the inclusion of dimension six operators touches all the ingredients of the calculation. 
For instance we have

\bq
\alpha(\mrM_{\sPZ}) = \alpha(0)\,\Bigl[ 1 - \Delta \alpha_{\Pl} - \Delta \alpha_{\PQt} - \Delta \alpha^{(5)}_{\had}
\Bigr]^{-1}
\eq

Corrections due to leptons and to the top quark contain both $\mrdim = 4$ and $\mrdim = 6$ terms; the hadronic part
is taken, as usual from data.

It is important to understand that the inclusion of $\mrdim = 6$ only in tree{-}diagrams introduces blind directions
in the space of Wilson coefficients. The inclusion of loops partially removes the degeneracy.

\section{Numerical results \label{NR}}

For the masses we use the values quoted by the PDG and define two scenarios corresponding to 

\begin{itemize}

\item[S1] $\Lambda = 1\UTeV$ with values of the (renormalized) 
          Wilson coefficients $\in\,[-0.1\,,\,+0.1]$. 

\item[S2] $\Lambda = 2\UTeV$ with values of the (renormalized) 
          Wilson coefficients $\in\,[-0.1\,,\,+0.1]$. 

\end{itemize}

We start with the
observation that $\alpha(\mrM_{\sPZ})$ in the SMEFT differs from the SM value by less than a permille.  

\paragraph{SMEFT and $\mra_{\PGm}$} \hspace{0pt} \\
The accepted theoretical value for $\mra_{\PGm}$ is $0.00116591810\,(43)$~\cite{Aoyama_2020}; 
the new experimental world{-}average results 
today is $0.00116592061\,(41)$~\cite{Muong-2:2006rrc} with a difference of $251\,\times\,10^{-11}$. 
Given the one{-}loop EW contribution~\cite{Bardeen:1972vi}

\bq
\mra^{\myEW}_{\PGm}\,\mid_{\mathrm{one{-}loop}} =
\frac{\mrG_{\mrF}\,M^2_{\PGm}}{24\,\sqrt{2}\,\pi^2}\,\Bigl[ 5 + \bigl(1 - 4\,\sin^2\theta_{\sPW}\bigr)^2 \Bigr]
\eq

where $\sin^2_{\sPW} = 0.22301$, we obtain
$\mra^{\myEW}_{\PGm} = 194.8\,\times\,10^{-11}$ at one loop (higher order EW corrections bring this value to
$153.6\,\times\,10^{-11}$~\cite{Czarnecki:2002nt}). Therefore the new experimental value requires 
deviations of $\ord{100}$ percent w.r.t. the SM which are obviously difficult to reach in the context of the SMEFT.
For instance, with $\Lambda = 1\UTeV$ and Wilson coefficients $\in\,[-0.5\,,\,+0.5]$ we can reach a $50\%$ 
deviations but only in the corners of the space of Wilson coefficients. Note that at tree level $\mra_{\PGm}$
depends only on $\alW$ and $\alB$ (Wilson coefficients in the Warsaw basis).

We define 

\bq
\Delta \mra_{\PGm} = \frac{\mra^{(6)}_{\PGm}}{\mra^{(4)}_{\PGm}} - 1 , \qquad
\mra^{(4)}_{\PGm} = \mra^{\myEW}_{\PGm}\,\mid_{\mathrm{one loop}} .
\eq

%In Fig.~\ref{devfig1} 
In \hyperref[devfig1]{Fig.(\ref*{devfig1})}
we show the pdf for $\Delta \mra_{\PGm}$. Left figure refers to S1, right figure to S2. We
have also produced results for $\Lambda = 1\UTeV$ and Wilson coefficients $\in\,[-0.5\,,\,+0.5]$. The result
is shown in 
\hyperref[devfig2]{Fig.(\ref*{devfig2})}.
%Fig.~\ref{devfig2}. 

\begin{figure}[t]
   \centering
   \vspace{-3.cm}
\includegraphics[height=7.cm,width=7.cm]{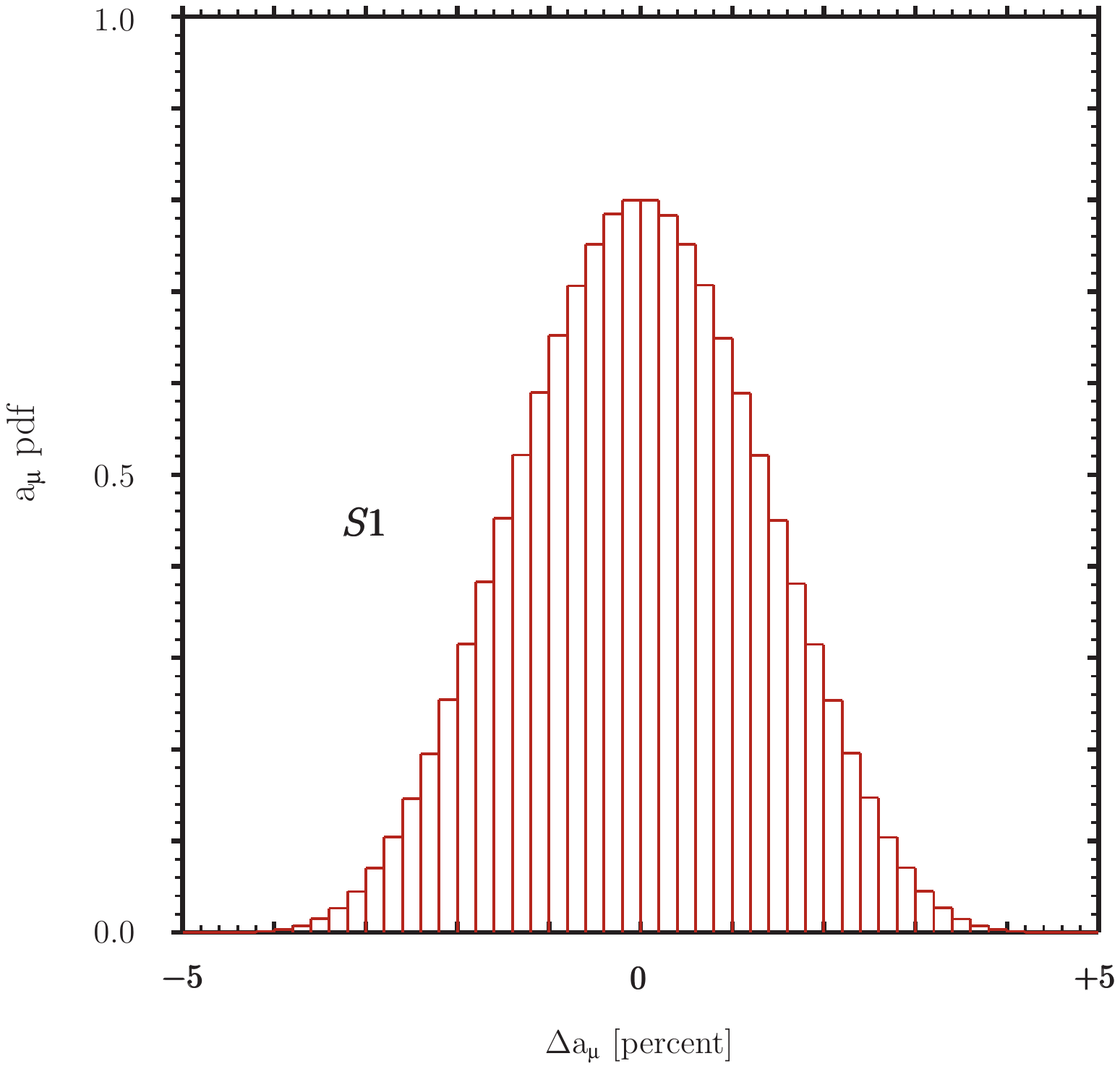}
\includegraphics[height=7.cm,width=7.cm]{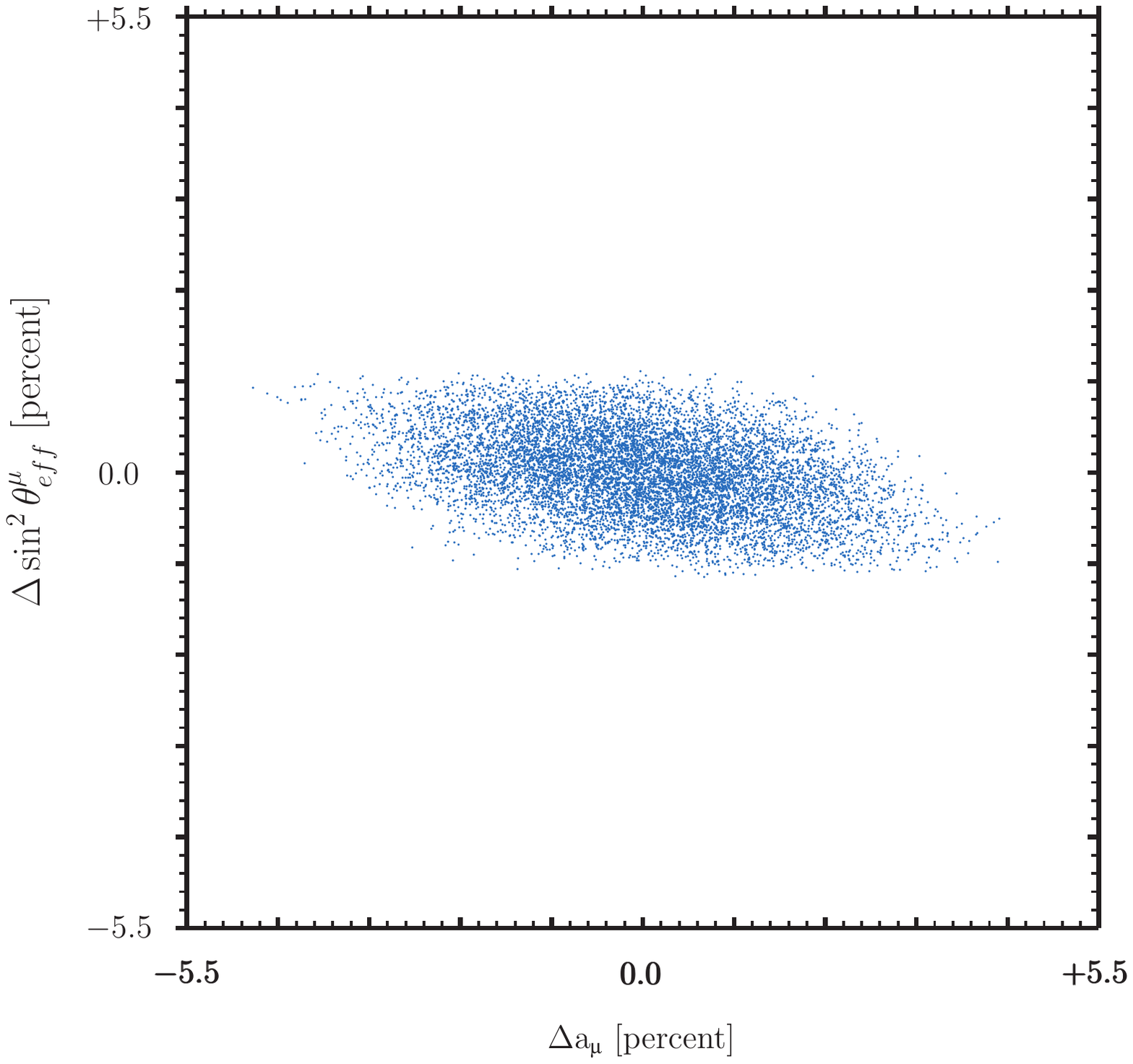}
%\vspace{-1.cm}
\vspace{-1.5cm}
\caption[]{The pdf for $\Delta \sin^2\theta^{\PGm}_{\eff}$ in scenario S1 (left figure). A scattered plot displaying the
relationship between $\Delta \mra_{\PGm}$ and $\Delta \sin^2\theta^{\PGm}_{\eff}$ (right figure).}
\label{devfig3}
\end{figure}

We have also investigated a scenario where $\Lambda = 1\UTeV$, Wilson coefficients $\in\,[ -1\,,\,+1 ]$ and where
we discard points in the space of Wilson coefficients where $\mid \Delta \sin^2\theta^{\PGm}_{\eff} \mid > 10^{-3}$. 
In the resulting distribution the largest fraction of $\Delta \mra_{\PGm}$ deviations is compatible with zero.
From the histograms we derive an approximate value for the standard deviation of the pdf: they are
$\sigma = 0.357$ for S2, $\sigma = 1.428$ for S1 and $\sigma = 7.142$ for the last scenario. 
For the sake of completeness we have repeated the calculation assuming the all Wilson coefficients are positive; 
the result shows a mean $\mu = 1.414$ with $\sigma = 1.429$.

By looking at the data we can ask ``what happens when we put constraints on the space of Wilson coefficients''? 
The situation seems to be the following: there is one combination of Wilson coefficients (in the Warsaw basis),
$\alWB = \sin\theta_{\sPW}\,\alW - \cos\theta_{\sPW}\,\alB$, which appears at tree level in the calculation of 
$\mra_{\PGm}$ but only at one{-}loop in the other pseudo{-}observables. Selecting large values for $\alWB$ while
putting to zero the remaining Wilson coefficients could do the job.
It is worth noting that according to \Bref{Einhorn:2013kja} $\alWB$ is the Wilson coefficient
of a Loop{-}Generated operator (containing field strengths), thus requiring a loop suppression factor.
The relevance of $\alW, \alB$ (called 
$\mra_{\Pe\,\scriptscriptstyle{\PW,\PB}}$ in \Bref{Grzadkowski:2010es}) becomes clear when we observe that the corresponding 
operators contain $\sigma^{\mu\nu}$.

To give an example, suppose that we use the SMEFT at the tree level; let us define

\bq
\aplV = \mra^{(3)}_{\upphi\,\Pl} - \mra^{(1)}_{\upphi\,\Pl} - \apl , \quad
\aplA = \mra^{(3)}_{\upphi\,\Pl} - \mra^{(1)}_{\upphi\,\Pl} + \apl , 
\eq

where the $\mra$ are Wilson coefficients in the Warsaw basis. We can derive both of them in terms of
other Wilson coefficients by asking zero deviation in the vector and axial couplings of the $\PZ{-}\,$boson. Then we fix
$\alWB = \sin\theta_{\sPW}\,\alW - \cos\theta_{\sPW}\,\alB$ in order to reproduce the $\mra_{\PGm}$ deviation with
the rest of the Wilson coefficients left free. Always accepting to work at tree level, if we fix the free Wilson 
coefficients $\in\,[-0.1\,,\,+0.1]$ values of $\alWB/(16\,\pi^2) > 0.01$ are needed to reach $50\%$ deviations
for $\mra_{\PGm}$. 

Alternatively, we can proceed as follows: every observable can be decomposed as  

\bq
\mcO = \mcO^{(4)} + \frac{\mrg_6}{\sqrt{2}}\,\Bigl[ \delta^{(6)}_{\mcO} + \frac{\mrG_{\mrF}\,\mrM^2_{\sPW}}{\pi^2}\,
\Delta^{(6)}_{\mcO} \Bigr] .
\label{LOvsNLO}
\eq

The first term in \eqn{LOvsNLO} represents the tree{-}level contribution in the SMEFT while the second accounts for 
loops in SMEFT. We have set to zero the $\delta^{(6)}$ terms in the vector and axial couplings of the $\PZ$ boson,
which means two linear conditions among Wilson coefficients.

\begin{figure}[t]
   \centering
%   \vspace{-3.cm}
\includegraphics[height=7.cm,width=7.cm]{SMEFTdev_fig6.pdf}
\includegraphics[height=7.cm,width=7.cm]{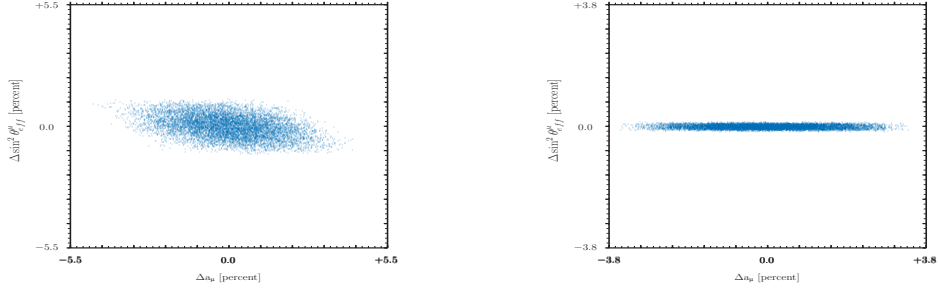}
\vspace{-2.cm}
\caption[]{A scattered plot displaying the relationship between $\Delta \mra_{\PGm}$ and 
$\Delta \sin^2\theta^{\PGm}_{\eff}$. Left figure is the same as in 
\hyperref[devfig3]{Fig.(\ref*{devfig3})}.
%Fig.~\ref{devfig3}. 
Right figure takes into account the constraint described in 
\hyperref[LOvsNLO]{Eq.(\ref*{LOvsNLO})}.
%\eqn{LOvsNLO}. 
}
\label{devfig8}
\end{figure}

The resulting scattered plot is shown in 
\hyperref[devfig8]{Fig.(\ref*{devfig8})}.
%Fig.~\ref{devfig8}. 
The interesting fact in comparing the left and right figures is
that the tree{-}level SMEFT correction is dominant but the one{-}loop SMEFT contribution 
is not negligible~\cite{Passarino:2016pzb,Freitas:2016iwx}. In the right panel of 
\hyperref[devfig2]{Fig.(\ref*{devfig2})} we have shown the (tiny) effect of changing the renormalization scale.

For $\Lambda = 1\UTeV$ and Wilson coefficients $\in\,[-0.1\,,\,+0.1]$, although we have reduced the deviations 
for $\sin^2\theta^{\PGm}_{\eff}$ at the level of $\pm 0.2\%$ we can only obtain deviations for $\mra_{\PGm}$ at the 
level of $\pm 4\%$.
 
\paragraph{SMEFT and $\PZ \to \PGmp\PGmm$} \hspace{0pt} \\

LEP data~\cite{Z-Pole} return $\sin^2\theta^{\Pl}_{\eff} = 0.23153 \pm 0.00016$ (the error is $0.07\,\%$). 
The natural comment is that it is
extremely difficult to evade this bound. Despite this evidence we should observe the following facts:
the LEP1 data were used to predict $\mrM_{\sPW} = 80.363 \pm 0.032\UGeV$ (we use $80.379\UGeV$) and the Higgs
boson mass was not an input parameter, it was obtained $\mrM_{\sPH} < 285\UGeV$ at $95\,\%$ C.L.

\begin{figure}[t]
   \centering
%   \vspace{3.cm}
\includegraphics[height=7.cm,width=7.cm]{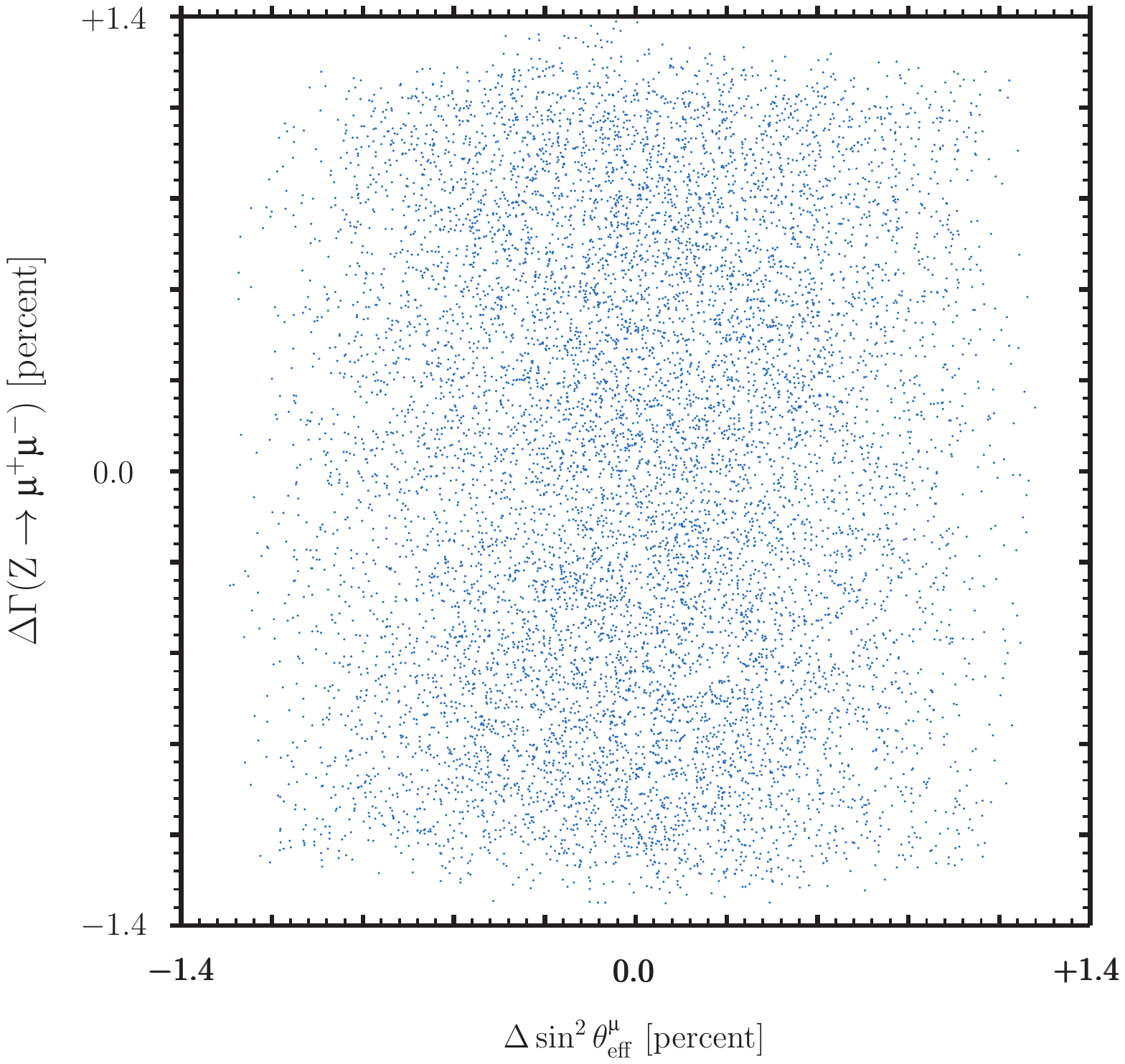}
\includegraphics[height=7.cm,width=7.cm]{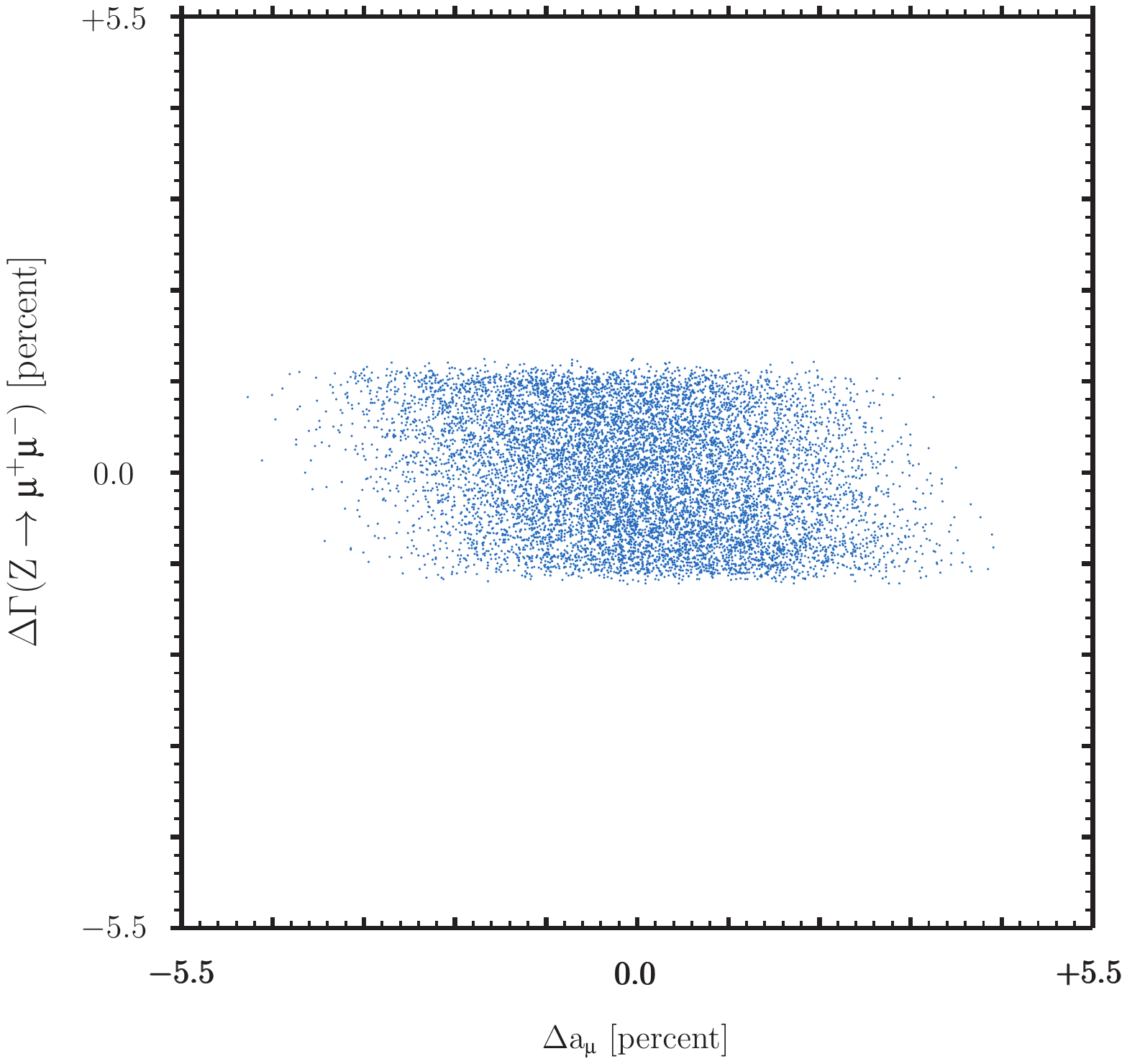}
%\vspace{-1.cm}
\vspace{-1.5cm}
\caption[]{A scattered plot displaying the relationship between $\Delta(\PZ \to \PGmp\PGmm)$ and
$\Delta \sin^2\theta^{\PGm}_{\eff}$ (left figure) or $\Delta \mra_{\PGm}$ (right figure).}
\label{devfig4}
\end{figure}

The pdf for $\sin^2\theta^{\PGm}_{\eff}$ is shown in 
\hyperref[devfig3]{Fig.(\ref*{devfig3})}
%Fig.~\ref{devfig3} 
for S1 and where we have defined

\bq
\Delta \sin^2\theta^{\PGm}_{\eff} = 
\frac{\sin^2\theta^{\PGm}_{\eff}\,\mid_{\mrdim = 6}}{\sin^2\theta^{\PGm}_{\eff}\,\mid_{\mrdim = 4}} - 1 .
\eq

We should remember that the corresponding experimental error is $0.07\%$. 
From the histogram we derive the following approximate deviation, $\sigma = 0.498$. 

\begin{itemize}
\item[\snitem]From the scattered plot
of 
%Fig.~\ref{devfig3} 
\hyperref[devfig3]{Fig.(\ref*{devfig3})}
we observe that, due to the low correlation, it is still possible to accommodate large
deviations for $\mra_{\PGm}$ while keeping (very) low deviations for $\sin^2\theta^{\mu}_{\eff}$. 
\end{itemize}

\paragraph{SMEFT and $\mrM_{\sPW}$} \hspace{0pt} \\

\begin{figure}[t]
   \centering
   \vspace{-2.cm}
\includegraphics[height=10.cm,width=10.cm]{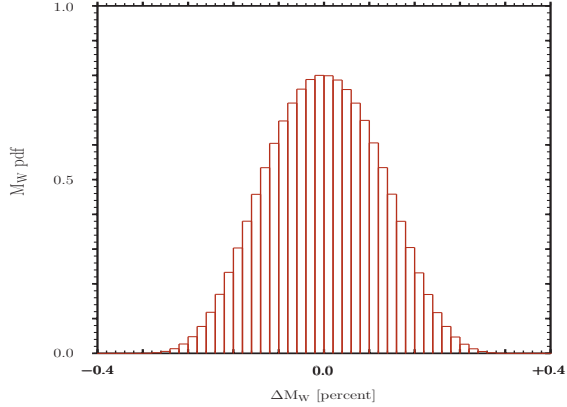}
\vspace{-2.5cm}
\caption[]{The pdf for $\Delta \mrM_{\sPW}$, scenario S1.}
\label{devfig5}
\end{figure}

The result from the CDF II detector is $\mrM_{\sPW} = 80.4335 \pm 0.0094 \UGeV$~\cite{CDF:2022hxs} to be 
compared with the previous world{-}average, $\mrM_{\sPW} = 80.379 \pm 0.012 \UGeV$~\cite{ParticleDataGroup:2022pth}. 
Therefore we require a deviation of ${+}\,0.068\%$. 
For fits see \Bref{Bagnaschi:2022whn}. Our result for the  $\Delta \mrM_{\sPW}$ pdf are shown in 
%Fig.~\ref{devfig5}.  
\hyperref[devfig5]{Fig.(\ref*{devfig5})}.
From the histogram we derive the following approximate deviation, $\sigma = 0.098$.

\paragraph{SMEFT and $\PH \to \PGg\PGg$, $\PH \to \PAQb\PQb$} \hspace{0pt} \\

\begin{figure}[t]
   \centering
%   \vspace{-3.cm}
\includegraphics[height=7.cm,width=7.cm]{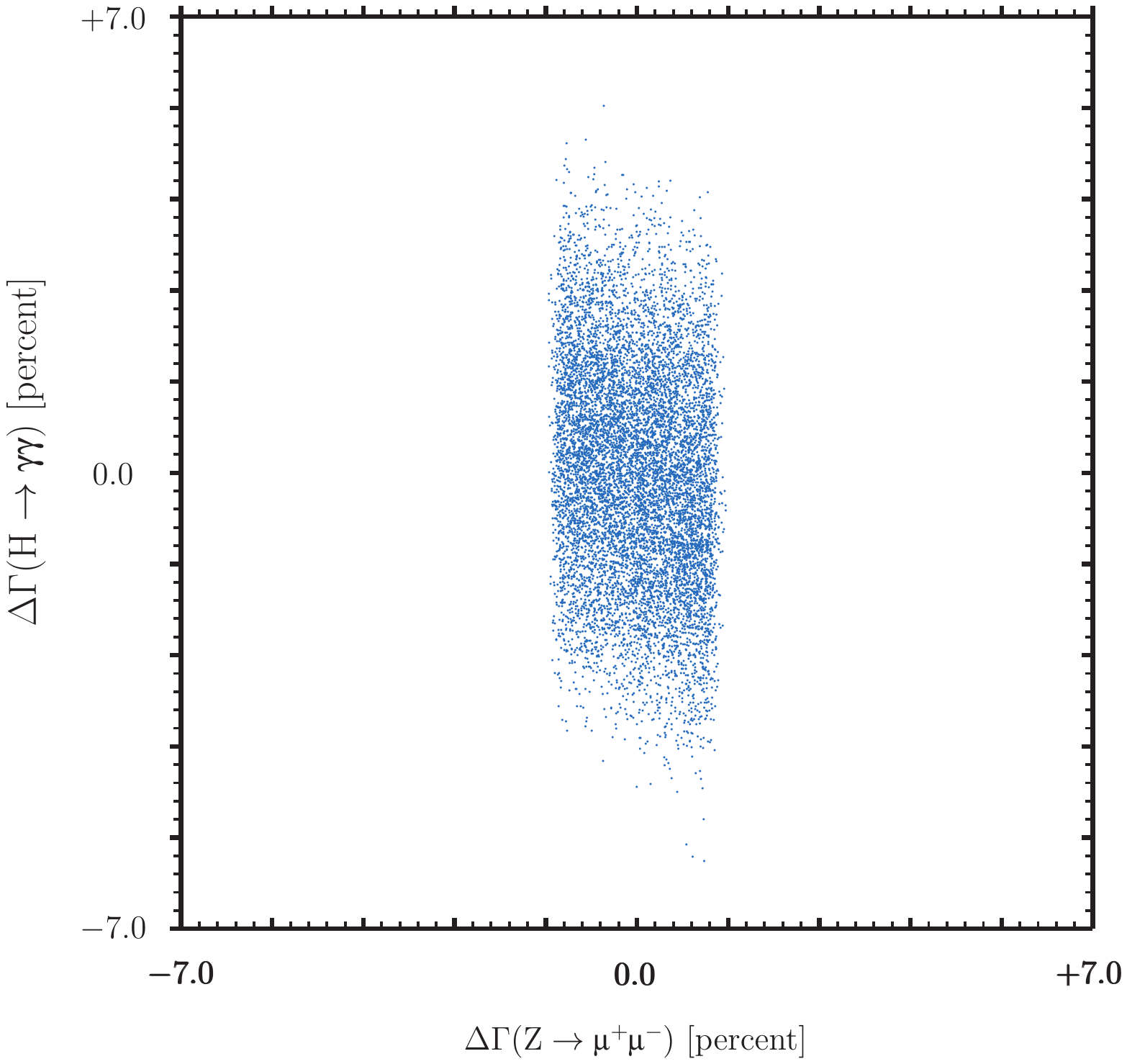}
\includegraphics[height=7.cm,width=7.cm]{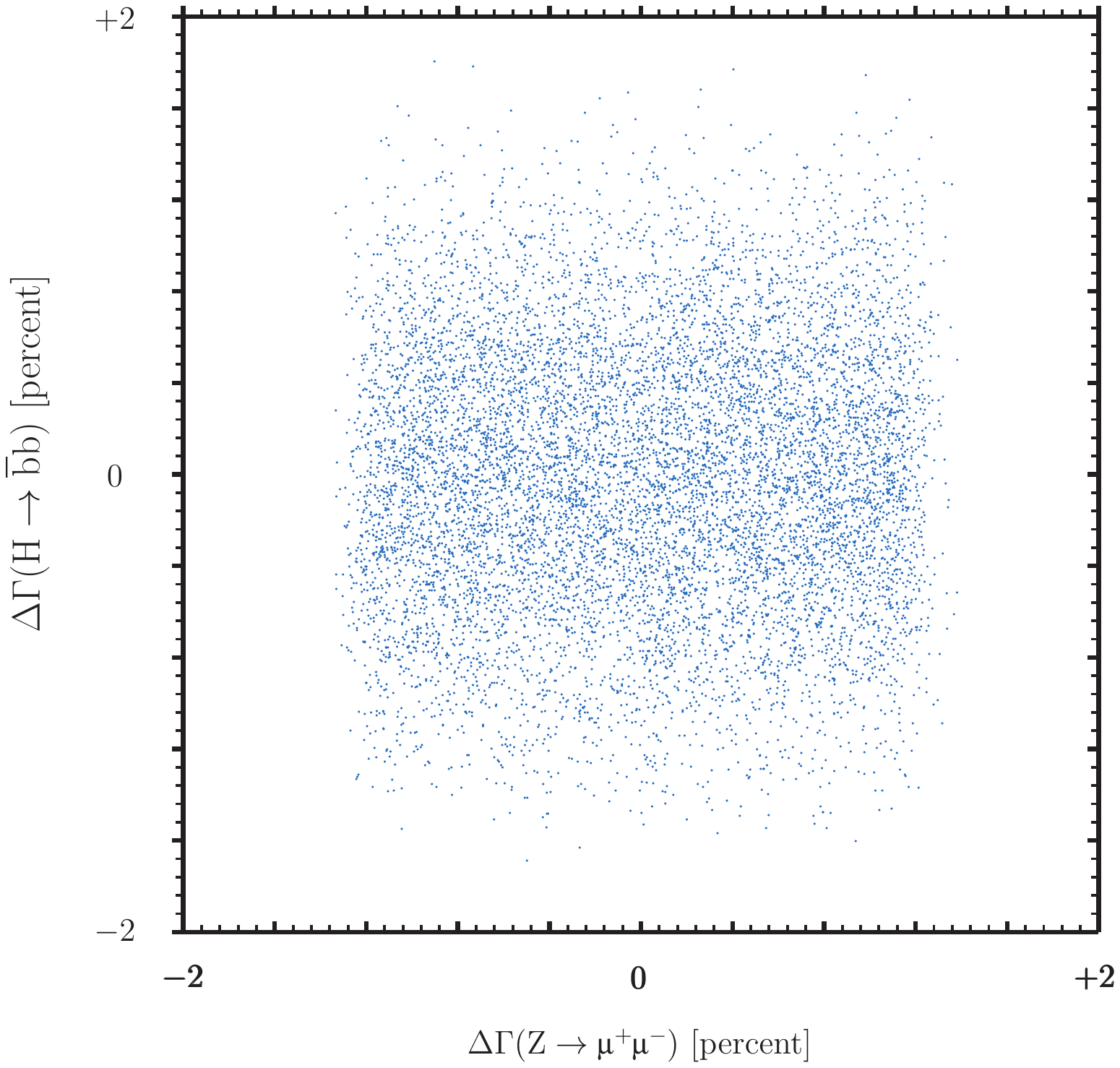}
\vspace{-1.cm}
\caption[]{A scattered plot displaying the relationship between $\Delta \Gamma(\PZ \to \PGmp\PGmm)$ and
$\Delta \Gamma(\PH \to \PGg\PGg)$ (left figure) or $\Delta \Gamma(\PH \to \PAQb\PQb)$ (right figure).}
\label{devfig6}
\end{figure}

In 
\hyperref[devfig6]{Fig.(\ref*{devfig6})}
%Fig.~\ref{devfig6} 
we show a scattered plot displaying the relationship between the set of ``data'' corresponding to
$\PZ \to \PGmp\PGmm$ and $\PH \to \PGg\PGg$.

%\paragraph{SMEFT and $\PH \to \PAQb\PQb$} \hspace{0pt} \\

In 
\hyperref[devfig6]{Fig.(\ref*{devfig6})}
%Fig.~\ref{devfig6} 
we show a scattered plot displaying the relationship between the set of ``data'' corresponding to
$\PZ \to \PGmp\PGmm$ and $\PH \to \PAQb\PQb$. For $\PH \to \PAQb\PQb$ we use the deconvoluted result (both QED and QCD).

From LHC Run 1 Higgs results (combined ATLAS and CMS results as reported in \Bref{Ellis:2018gqa}) 
the $\PGg\PGg$ decay (production mechanism $\Pg\Pg\PH$) the signal strength is $1.10^{+0.23}_{-0.22}$ while 
the $\PAQb\PQb$ decay ($\PV\PH$ production) is $1.0 \pm 0.5$ ($1.01 \pm 0.20$ for Run 2).

\paragraph{Inpact of QED corrections on $\PZ \to \PGmp\PGmm$} \hspace{0pt} \\

Let us consider QED corrections to the decay $\PZ \to \PGmp\PGmm$. After adding up virtual and real contributions
and defining the linear combination of Wilson coefficients, the final result for the decay width
is an IR/collinear-free quantity, both in the SM and in the SMEFT.
The result can be written as

\begin{figure}[t]
   \centering
%   \vspace{-3.cm}
\includegraphics[height=10.cm,width=10.cm]{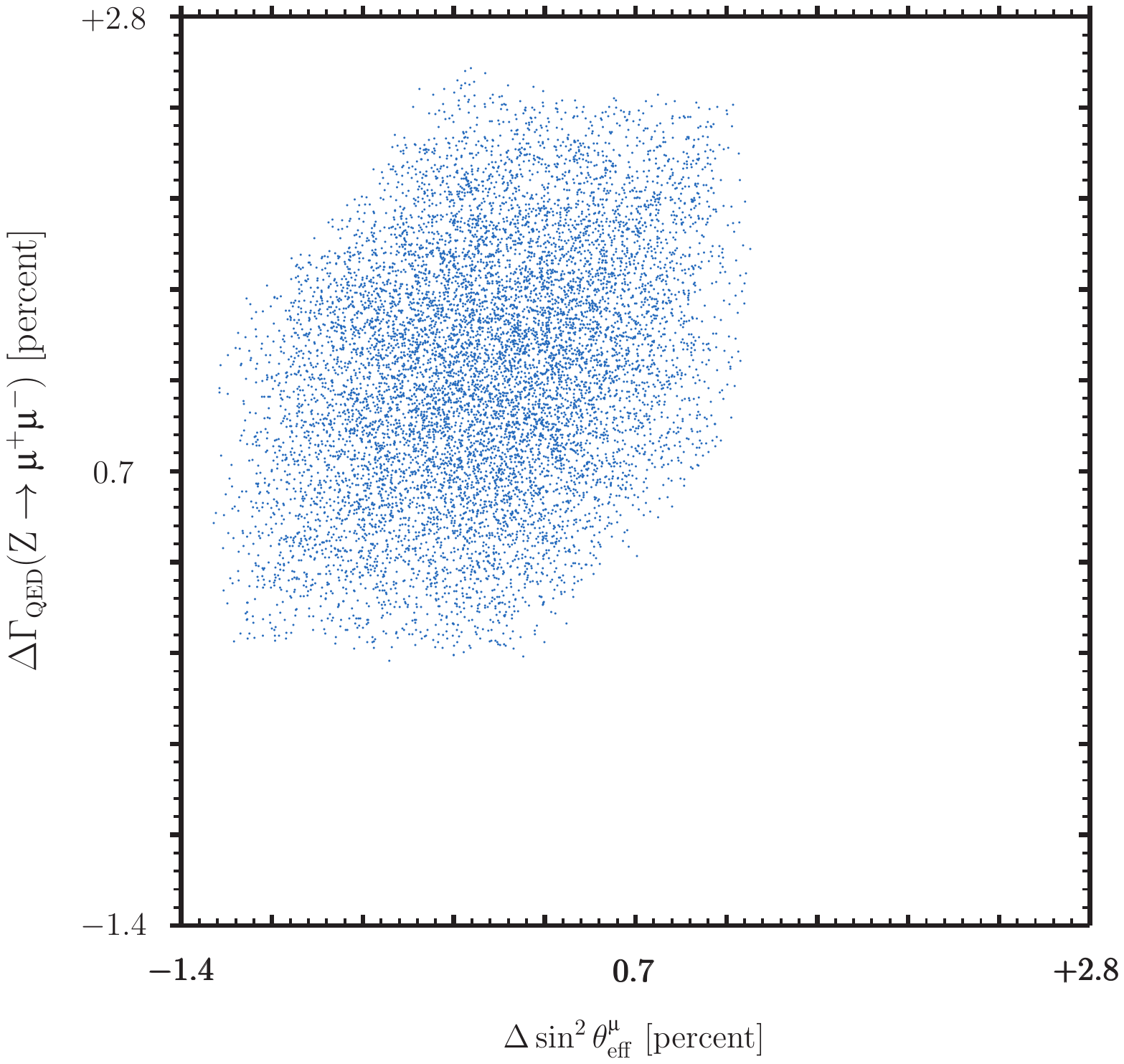}
%\vspace{-2.cm}
\vspace{-2.5cm}
\caption[]{A scattered plot displaying the relationship between $\Delta \Gamma_{\QED}(\PZ \to \PGmp\PGmm)$ and
$\Delta \sin^2\theta^{\PGm}_{\eff}$.}
\label{devfig7}
\end{figure}

\bq
\Gamma_{\QED}(\PZ \to \PGmp\PGmm) = \frac{3}{4}\,\frac{\alpha}{\pi}\,\Gamma_0\,\Bigl[
(1 + \mrv^2_{\Pl})\,\bigl(1 + \frac{\mrg_6}{\sqrt{2}}\,\delta^{(6)}_1 \bigr) + \frac{\mrg_6}{\sqrt{2}}\,\delta^{(6)}_2 
\Bigr] ,
\label{QED}
\eq

with $\mrv_{\Pl} = 1 - 4\,\sin^2\theta_{\sPW}$, showing a double (SM and SMEFT) factorization. The scattered plot 
displaying the relationship between $\Delta \sin^2\theta^{\PGm}_{\eff}$ and 
$\Delta \Gamma_{\QED}(\PZ \to \PGmp\PGmm)$ is shown in 
\hyperref[devfig7]{Fig.(\ref*{devfig7})}.
%Fig.~\ref{devfig7}.

This result is important, not only for extending IR/collinear finiteness to the SMEFT but also
because it shows that higher dimensional operators enter everywhere: signal, background and
radiation. The latter is particularly relevant when one wants to include (SM-deconvoluted)
EW precision observable constraints in a fit to Higgs data. Since LEP POs are (mostly) SM-
deconvoluted, the effect of $\mrdim = 6$ operators on the deconvolution procedure should be checked
carefully.

In 
\hyperref[QED]{Eq.(\ref*{QED})}
%\eqn{QED} 
$\Gamma_0$ is the LO width and the LEP definition is 

\bq
\Gamma^{\dec}(\PZ \to \PAl\Pl) = \frac{\Gamma(\PZ \to \PAl\Pl)}{1 + \frac{3}{4}\,\frac{\alpha(\mrM^2_{\sPZ})}{\pi}} .
\eq

Once again, fitting the deconvoluted pseudo-observable as reported at LEP with the SMEFT is not fully consistent.

\section{Conclusions \label{Conc}}

Without looking at the experimental data we have assumed 
a bottom{-}up approach described by the
SMEFT. Therefore, the randomly generated Wilson coefficients of the SMEFT (in the
Warsaw basis) are treated as pseudo{-}data and, for each observable, we
have computed the corresponding probability density function. 

The equiprobability bias (EB) is a tendency to believe that every process
in which randomness is involved corresponds to a fair distribution, with
equal probabilities for any possible outcome.
It has been shown that the EB is actually not the result of a conceptual
error about the definition of randomness.
On the contrary, the mathematical theory of randomness does imply uniformity~\cite{EB}.

Stated differently we consider a single ``supersystem'' where we identify the values of the Wilson
coefficients and implement the principle that equal ignorance should be represented by equiprobability.

Our goal has been to understand how large are the deviations from the SM once the SMEFT
scale ($\Lambda$) and the range of the Wilson coefficients are selected.
Theory{-}driven research~\cite{driven} focuses on identifying abstract constructs and the relationships
among them.

Our set of pseudo{-}observables includes $\mra_{\PGm}$, $\sin^2\theta^{\PGm}_{\eff}$, $\mrM_{\sPW}$,
$\Gamma(\PZ \to \PGmp\PGmm)$, $\Gamma(\PH \to \PGg\PGg)$ and $\Gamma(\PH \to \PAQb\PQb)$. 
The corresponding results can be compared with the experimental data to understand how easy or
difficult will be to explain (possible) SM{-}deviations in terms of the SMEFT language. In
particular, constraints in the space of Wilson coefficients have been introduced to 
explain large deviations for one observable and very small deviations for a second observable.
These constraints put you on the hedge of the equiprobable space of Wilson coefficients with
some of them taking ``very'' large values.

Theories can be changed by data or even invalidated by them.
We have an initial theory (the SM) and a theory of deviations (the SMEFT), and then
we add and absorb new data, altering the theories at each point.

\vspace{1.cm}
{\bf{Acknowledgments}}:
G. P. gratefully acknowledges a constructive correspondence with A.~David.
%--
\clearpage

%--
\appendix
%--
\section{Details on renormalization \label{det}}

Since a large part of the renormalization procedure depends on two{-}point functions we briefly summarize the
procedure.
Let $\PX$ be any boson field, the inverse $\PX$ propagator (with $p^2 = - \mrs$) is

\bq
- \mrs + \mrm^2_{\sPX} - \frac{\mrg^2}{16\,\pi^2}\,\Sigma_{\sPX}(\mrs) ,
\eq

where $\mrm_{\sPX}$ is the bare $\PX$ mass. We introduce CTs, \ie

\bq
\mrm_{\sPX} = \mrm^{\ren}_{\sPX}\,\Bigl\{ 1 + \frac{\mrg^2}{16\,\pi^2}\,\Bigl[
\delta^{(4)}_{\mrm_{\PX}} + \frac{\mrg_6}{\sqrt{2}}\,\delta^{(6)}_{\mrm_{\PX}} \Bigr] \Bigr\} ,
\eq

and remove $\mrs{-}$independent UV poles. Next we write the renormalization equation at $\mrs = \mrM^2_{\sPX}$ where
$\mrM_{\sPX}$ is the physical (on{-}shell) mass.

\bq
\Sigma_{\sPX}(\mrs) = \Sigma_{\sPX}(\mrM^2_{\sPX}) + ( \mrs - \mrM^2_{\sPX} )\,\Sigma^{\prime}_{\sPX}(\mrm^2_{\sPX})
+ \mathrm{rest} .
\eq

Introducing now

\bq
\mrm^{\ren}_{\sPX} = \mrM_{\sPX}\,\Bigl\{ 1 + \frac{\mrg^2}{16\,\pi^2}\,\Bigl[
\Delta^{(4)}_{\mrm_{\PX}} + \frac{\mrg_6}{\sqrt{2}}\,\Delta^{(6)}_{\mrm_{\PX}} \Bigr] \Bigr\} ,
\eq

we fix the new CTS such that

\bq
\mrM^2_{\sPX} = \mrm^2_{\sPX} - \frac{\mrg^2}{16\,\pi^2}\,\Re\,\Sigma_{\sPX}(\mrM^2_{\sPX}) ,
\eq

and derive the corresponding wave{-}function factors. For fermions the procedure requires the introduction of
$1 \pm \gamma^5$ projectors and will not be repeated here.

The inclusion of vertices and boxes in the amplitude (after the introduction of the wave{-}function factors 
for the external legs) is such that

\begin{itemize}

\item[\snitem] for the SM ($\mrdim = 4$) the amplitudes are UV finite,

\item[\snitem] for the SMEFT ($\mrdim = 6$) we have to introduce a mixing

\bq
\mra_i = \mrZ_{ij}\,\mra^{\ren}_j , \qquad
\mrZ_{ij} = \delta_{ij} + \frac{\mrg^2}{16\,\pi^2}\,\Bigl[
\delta \mrZ^{(4)}_{ij} + \frac{\mrg_6}{\sqrt{2}}\,\delta \mrZ^{(6)}_{ij} \Bigr] ,
\eq

where $\mra_i$ are Wilson coefficients.

\end{itemize}

%--
\clearpage
%--
%\bibliographystyle{atlasnote}
\bibliographystyle{elsarticle-num}
\bibliography{SMEFTdev}{}

%===
\end{document}